# Confirmation that T cell receptor activation is digital, not analog. Hence individual T cell cytolytic capacity is independent of initial stimulation strength

**Abstract.** Whether a sub-optimum lymphocyte stimulus is achieved by lowering ligand concentration, or decreasing its affinity, initial TCR activation responses are likely to be digital. We now know for cultured peripheral blood mononuclear cells (PBMCs), when activated by plant lectins such as concanavalin-A (Con-A), that the responding cells are T cells, and that early transcriptional and metabolic changes closely resemble those found with T cells responding to specific peptides complexed to MHC proteins (pMHC). Robbins long ago showed that lectin-activated PBMC responses were digital. This has now been confirmed for pMHC-activated T cells. Hence, in general, responses of individual T cells are independent of initial signal strength.

___

Richard et al. have recently asked whether providing T cells with a sub-optimal early activating stimulus will cause all cells to respond sub-optimally? Alternatively, as they proposed, will some cells respond fully, while the rest are responseless[1]? As formulated by Balyan et al., with respect to the commitment of an individual cell, is the response analog (continuous) or digital (all-or-none)[2]? If there is a suboptimum stimulus, the potency of which can relate to ligand affinity and/or dosage, will *all* cells respond less fully, or will a full response be manifest in only a proportion of the cells? To examine this Richard et al. applied emerging single cell technologies to a T cell population engineered to possess identical T cell receptors (TCRs) of uniform affinity for a specific peptide ligand that could be added directly to cell cultures.

    However, the experiment could also have been carried out with a T cell population bearing TCRs that are not uniform with respect to affinity for peptide ligands. In this case one would need a general, polyclonal, ligand that did not influence the peptide-reactive sites of TCRs but was uniformly reactive with certain sites that are common to all TCRs. Furthermore, one would need to show that responses to activation are identical (i.e. have shared activation machinery)

whether brought about in this polyclonal fashion, or monoclonally (as with Richard et al.). Remarkably, the result of the polyclonal experiment became available in 1963, prior to clarification of the distinction between T and B cells with their corresponding receptors. Nevertheless, it can now be recognized as supporting the digital TCR proposal of Richard et al.[3].

Shortly after Nowell's discovery for cultured peripheral blood mononuclear cells (PBMC) of the polyclonal activating power of certain plant lectins, Robbins showed by simple microscopy that sub-optimal lectin concentrations activated only a proportion of the cells, leaving other potentially activatable cells unaffected[3,4]. For many reasons this appears highly relevant to the present study. Plant lectins, such as phytohaemagglutinin (PHA) and concanavalin-A (Con-A), activate TCRs independently of antigen[5]. The Con-A tetramer, for example, is a mannose-binding lectin (MBL) that binds TCR-associated N-linked glycans. This activates the same intracellular enzymic cascades as are activated in specific pMHC-activated T cells and results in early transient expression of mRNAs for transcription factors (e.g. Egr1, FosB), an enhancement of glycolysis, and cell proliferation[6]. The mRNAs for Egr1 and FosB were among those cDNAs cloned in the 1980s from human PBMC based on their differential expression in the first two hours of culture with lectin[7].

As Richard et al. show, Egr1 and FosB mRNAs are also expressed early in peptide-activated monoclonal mouse T cells. However, despite uniformities of both the activating stimulus and the cells, there was considerable variation, which was attributed to "environmental fluctuations and intercellular variability." Furthermore, there was doubt as to whether stimuli were TCR-dependent or independent. Clarification of the results required elegant mathematical procedures such as diffusion pseudotime analysis that, nevertheless, did not eliminate baseline stimulations in cells deemed as "resting." In contrast, the Con-A experiments were carried out with freshly explanted PBMC that were "rested" for a day under culture conditions before challenging with ligand. Without this "rest" period, immediate-early gene transcription appeared to be provoked by the procedures involved in cell purification and culture establishment[8-11]. Figures 1 and 2 show activation kinetics close to those reported by Richard et al., but very low baselines in unstimulated cultures. In contrast, as particularly evident with FosB mRNA, Richard et al. have high baseline stimulations in non-peptide treated cultures.

In conclusion, whether a sub-optimum stimulus is achieved by lowering ligand concentration, or decreasing its affinity, initial TCR activation responses are likely to be digital. However, as Richard et al. acknowledge, the digital/analogue issue is contentious, so supporting evidence from other sources should be of value. We now know for cultures of lectin-treated PBMCs that the responding cells are T cells, that they respond digitally by way of their TCRs, and that subsequent metabolic changes closely resemble those found with T cells responding to specific pMHCs. The likely physiological relevance of Con-A studies is underscored by the observation that, like endogenous animal MBLs, Con-A can activate complement by way of what is now known as the lectin complement pathway[12]. Of special importance when cultures contain unheated serum is the possibility that a ligand may bring about complement-dependent inhibitory effects (perhaps simulating immunological tolerance); this may be missed when prior to examination there is gating to exclude dead cells. Unlike the peptide ligand studies of Richard et al., for lectin ligands simply grouping cells by sampling time can "achieve the fine resolution needed to understand coordination of gene expression," and much more.

**Donald R. Forsdyke**

*Department of Biomedical and Molecular Sciences, Queen's University, Kingston, Ontario, Canada K7L3N6*

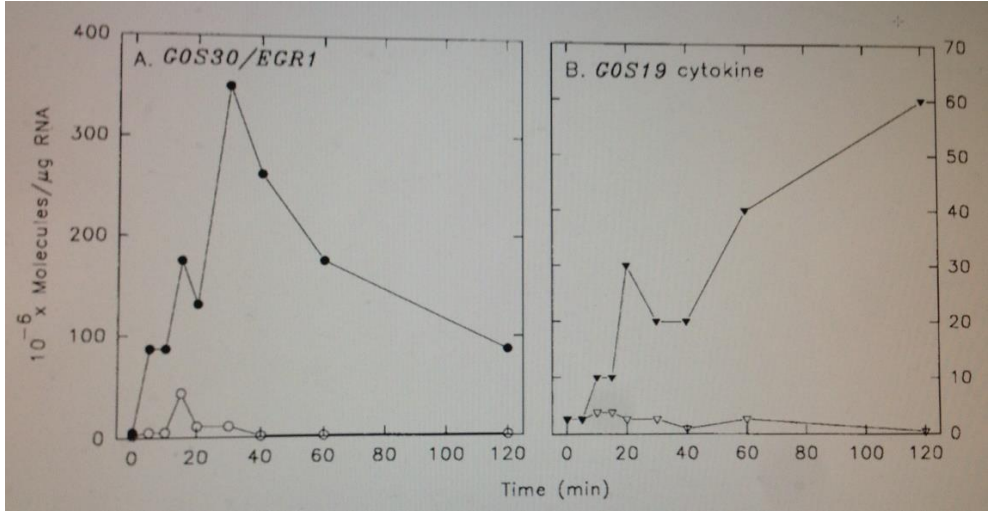

**Fig. 1 | Increases in mRNAs in Con-A-activated human PBMC (filled symbols) during first two hours of culture. A**, RNA for the transcriptional activator Egr1 (G0S30). **B**, RNA for the chemokine CCL3 (MIP1α, G0S19). Freshly prepared cells were "rested" in culture for a day prior to the addition of Con-A at 0 min.. For details see reference 8.

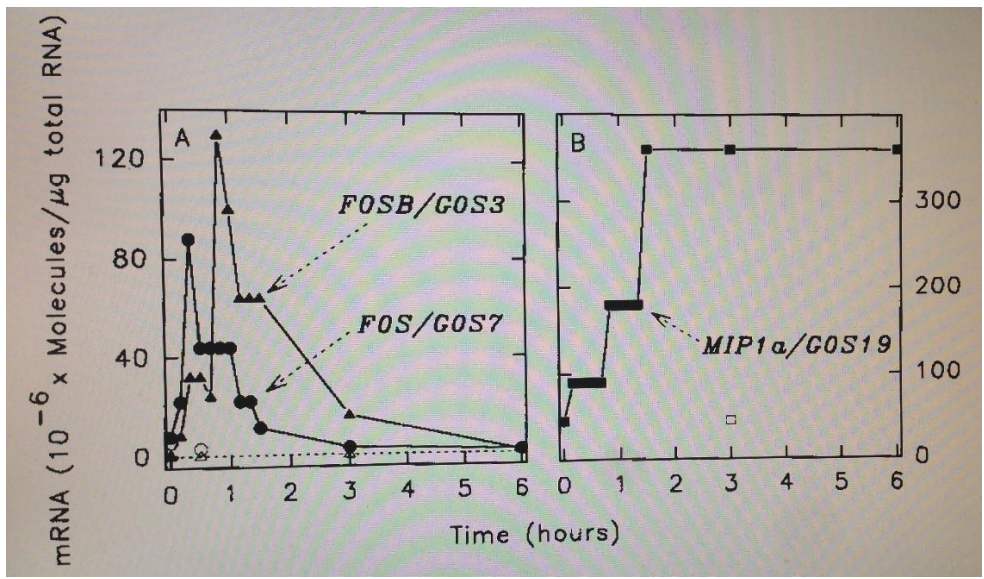

**Fig. 2 | Increases in mRNAs in Con-A-activated human PBMC (filled symbols) during first six hours of culture. A**, RNA for the transcriptional activators Fosb (G0S3) and Fos (GOS7). **B**, RNA for the chemokine CCL3 (MIP1α, G0S19). Freshy prepared cells were "rested" in culture for a day prior to addition of Con-A at 0 min.. For details see reference 9.